\documentclass[aps,prb,twocolumn,showpacs,amsmath,amssymb]{revtex4}

\usepackage{graphicx}
\usepackage{natbib}
\usepackage{color}
\usepackage{txfonts}

\newcommand{\kets}[1]{| \, #1 \rangle}

\begin{document}

\title{Josephson effect in normal and ferromagnetic topological
  insulator planar, step and edge junctions}

\author{Jennifer Nussbaum}
\author{Thomas L. Schmidt}
\author{Christoph Bruder}
\author{Rakesh P. Tiwari}
\email[Email address: ] {rakesh.tiwari@unibas.ch}
\affiliation{Department of Physics, University of Basel,
  Klingelbergstrasse 82, CH-4056 Basel, Switzerland}
\date{\today}

\begin{abstract}
  We investigate Josephson junctions on the surface of a
  three-dimensional topological insulator in planar, step, and edge
  geometries.  The elliptical nature of the Dirac cone representing
  the side surface states of the topological insulator results in a
  scaling factor in the Josephson current in a step junction as
  compared to the planar junction. In edge junctions, the contribution
  of the Andreev bound states to the Josephson current vanishes due to
  spin-momentum locking of the surface states. Furthermore, we
  consider a junction with a ferromagnetic insulator
  between the superconducting regions. In these ferromagnetic
  junctions, we find an anomalous finite Josephson current at zero
  phase difference if the magnetization is pointing along the junction
  (and perpendicular to the Josephson current).  An out-of-plane
  magnetization with respect to the central region of the junction
  opens up an exchange gap and leads to a non-monotonic behavior of
  the critical Josephson current for sufficiently large magnetization
  as the chemical potential increases.
\end{abstract}

\maketitle
\section{\label{sec:Introduction}Introduction}

Topological insulators are states of quantum matter whose electronic
structure cannot be adiabatically connected to conventional insulators
and semiconductors. They are characterized by an insulating gap in the
bulk and gapless edge states (in case of a two-dimensional (2D) topological
insulator) or surface states (in case of a three-dimensional (3D)
system) which are protected by time-reversal (TR) symmetry against
disorder and other perturbations that respect TR
symmetry.\cite{Kane2005a,Kane2005b,Fu2007a,Moore2007,Roy2009,Hasan2010,Qi2011}
The theoretical prediction of symmetry-protected edge states in HgTe
quantum wells\cite{Bernevig2006} led to the experimental
demonstration of HgTe quantum wells being 2D topological
insulators.\cite{Konig2007} Similarly, the theoretical prediction of
Bi$_{1-x}$Sb$_{x}$ being a 3D topological insulator\cite{Fu2007}
soon led to the experimental demonstration of 2D topological surface
states in Bi$_{0.9}$Sb$_{0.1}$.\cite{Hsieh2008} More compounds were
predicted to be 3D topological insulators using first-principles
electronic structure calculations, which include Sb$_2$Te$_3$,
Bi$_2$Te$_3$ and Bi$_2$Se$_3$.\cite{Zhang2009} The surface states of
these topological insulators were identified using angle-resolved
photo-emission spectroscopy\cite{Xia2009, Hsieh2009, Chen2009} and
scanning tunneling
microscopy.\cite{Zhang2009_2,Alpichshev2010}


In this article we consider 3D topological insulators with symmetry
protected 2D surface states. A simple low-energy effective model can
be shown to describe the topological insulators
Bi$_2$Se$_3$, Bi$_2$Te$_3$, and Sb$_2$Te$_3$ with a single Dirac cone
on the surface.\cite{Zhang2009} The topological insulator Bi$_2$Se$_3$ exhibits a circular
Dirac cone on the surface perpendicular to the
three-fold rotation symmetry axis (which we will call the top
surface), and an elliptical Dirac cone on the
side surfaces.\cite{Zhang2012,Alos-Palop2013}
In the case of Bi$_2$Se$_3$, this ellipticity suppresses the
conductance in a nanostep junction.\cite{Alos-Palop2013}

Recently, the consequences of induced
super\-con\-duc\-tivity\cite{Stanescu2010,Wang2012} and ferromagnetism at
the surface of topological insulators have attracted a great deal of
attention. In Refs.~[\onlinecite{Linder2010,Tanaka2009}] transport
properties of planar topological ferromagnetic junctions were
studied. The authors calculated the Josephson current of such
superconducting--ferromagnetic--superconducting (SFS) junctions and
found an anomalous current-phase relation for a magnetization pointing
in the direction of transport. This magnetization leads to a shift of
the phase difference in the Josephson junction, such that a finite
Josephson current is possible even at a phase difference $\phi = 0$.
The study in Refs.~[\onlinecite{Linder2010,Tanaka2009}] is based on
a Dirac-type surface Hamiltonian
\begin{align}\label{eq:Dirac}
    H_{\rm Dirac} = \hbar v_F\left(\sigma_x k_x + \sigma_y k_y\right),
\end{align}
where $v_F$ denotes the Fermi velocity.

\begin{figure}[t]
\includegraphics[width=0.89\columnwidth]{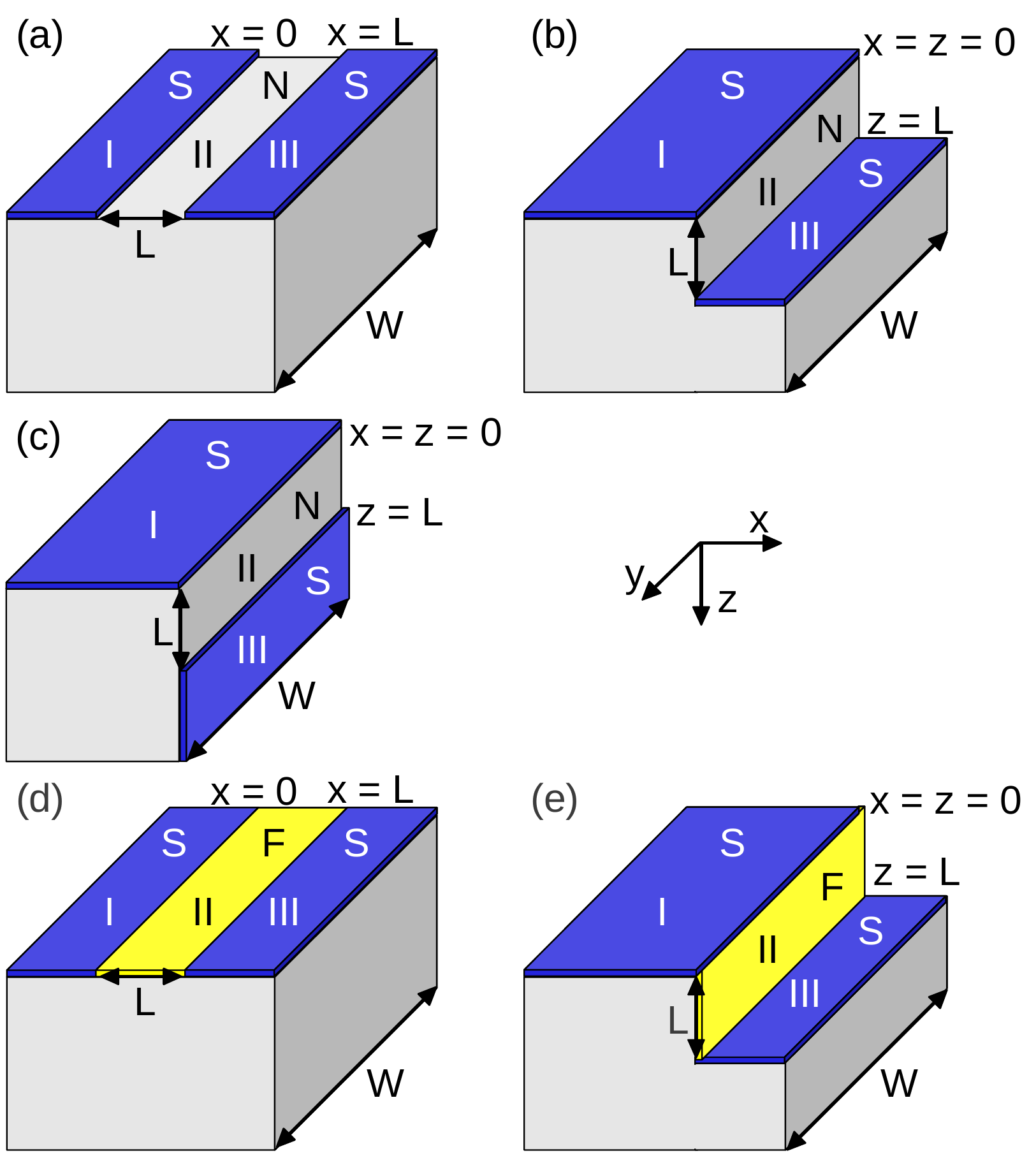}
\caption{\label{fig:bulk}(Color online) (a) Planar, (b) step and (c) edge topological
  insulator junction and ferromagnetic (d) planar and (e) step
  junction. The S-planes are the superconductors inducing an effective p-wave
  superconductivity in the surface of the topological insulator
  (regions I and III). The weak link (i.e., the central region) of the
  junction in (a), (b) and (c) is the pristine normal-conducting (N)
  surface of the topological insulator (region II). In (d) and (e),
  region II is covered by a ferromagnetic insulator with
  magnetization $\mathbf{M}$ that induces an exchange coupling
  in the surface states.}
\end{figure}

In the present manuscript we investigate the impact of the elliptical
nature of the Dirac cone on the proximity induced superconductivity in
the surface states of a 3D topological insulator. We quantify our
study by calculating the Josephson effect in superconducting--normal--superconducting (SNS) and SFS junctions on the surface of a 3D
topological insulator involving two different side surfaces. We show
that by measuring the critical current of these junctions, we can
quantify the ellipticity of the Dirac cone, providing information
about the bulk band structure and symmetry properties of these
topological insulators. For concreteness, we use the effective
low-energy model of the topological insulator Bi$_2$Se$_3$. First, we
analyze the influence of the ellipticity of the Dirac cone on the
different surfaces by evaluating the Josephson current in SNS planar,
step and edge junctions as shown in Fig.~\ref{fig:bulk} (a), (b) and
(c). Afterwards, we calculate the Josephson effect in the SFS planar
and step junctions shown in Fig.~\ref{fig:bulk} (d) and (e), and
analyze the dependence of the Josephson current on the direction of
magnetization.  Our results complement those of
Refs.~[\onlinecite{Linder2010,Tanaka2009}] because the surface states
of the Bi$_2$Se$_3$ topological insulator are governed by a
Rashba-type Hamiltonian of the form
\begin{align}\label{eq:Rashba}
H_{\rm Rashba} = \hbar v_F\left(\sigma_xk_y-\sigma_yk_x\right),
\end{align}
in contrast to the Dirac Hamiltonian shown in Eq.~(\ref{eq:Dirac}).


\section{\label{sec:Model}Model}
\subsection{\label{sec:Hamiltonians}Hamiltonians and wave functions}
Our goal is to calculate the Josephson effect for SNS planar, step and edge
junctions and SFS planar and step junctions. The calculations are done
for the 3D topological insulator Bi$_2$Se$_3$ and can be adapted to
any topological insulator whose surface states are described by a
Rashba-type Hamiltonian. The effective low-energy Hamiltonian for
Bi$_2$Se$_3$ in the basis of four hybridized states of Se and Bi
$p_z-$orbitals denoted as $\kets{P1^+_z \uparrow }$,
$\kets{P2^-_z\uparrow}$, $\kets{P1^+_z \downarrow}$,
$\kets{P2^-_z\downarrow }$ can be written as \cite{Zhang2009}
\begin{equation}
H(\mathbf{k}) = \varepsilon_0(\mathbf{k}) \mathbb{I}_{4 \times 4}+
\begin{pmatrix} M(\mathbf{k}) & A_1k_z & 0 & A_2k_- \\ A_1 k_z & -M(\mathbf{k}) & A_2k_- & 0 \\0 & A_2k_+ & M(\mathbf{k}) & - A_1k_z \\ A_2 k_+ &0&-A_1k_z &-M(\mathbf{k}) \end{pmatrix},
\end{equation}
where $k_\pm = k_x \pm i k_y$, $\varepsilon_0(\mathbf{k}) = C +
D_1k_z^2 +D_2k_+k_-$, $M(\mathbf{k}) = M - B_1 k_z^2 - B_2 k_+k_-$,
and $k_+k_- = k_x^2 + k_y^2$. The parameters $A_1, A_2, B_1, B_2, C, D_1, D_2,$
and $M$ can be determined by fitting the energy spectrum of this effective
Hamiltonian with that of the \textit{ab initio} calculations, see Ref.~[\onlinecite{Zhang2009}].
In the basis states, $\uparrow$ ($\downarrow$) stands
for spin up (down) and $+$ ($-$) stands for even (odd) parity. There
exists a straightforward procedure to obtain the effective Hamiltonian
describing the surface states.\cite{Qi2011} The effective surface Hamiltonian
for the $x-y$ plane of the topological insulator is
then given by \cite{Qi2011,Fu2009}
\begin{equation}
 H^{xy} = \varepsilon_0^{xy} + \hbar v_F^{xy} ( \sigma_x k_y - \sigma_y k_x),
\end{equation}
where $ \varepsilon_0^{xy} = C + (D_1/B_1)M$ is the Dirac point
energy, $v_F^{xy} = A_2 \sqrt{1-(D_1/B_1)^2} /\hbar$ represents the
Fermi velocity in the $x-y$ plane, and $\sigma_{x,y,z}$ denote
the Pauli matrices.

In contrast, the $y-z$ plane is described by the surface Hamiltonian
\begin{equation}
 H^{yz} = \varepsilon_0^{yz} + \hbar v_F^{yz} ( \sigma_y \eta k_z - \sigma_z k_y),\label{yzsurface}
\end{equation}
with the Dirac point energy $ \varepsilon_0^{yz} = C + (D_2/B_2)M$ and
the Fermi velocity $ v_F^{yz} = A_2 \sqrt{1-(D_2/B_2)^2} /\hbar$.  The
prefactor $\eta := A_1/A_2$ in front of $k_z$ in $H^{yz}$ is a
manifestation of the elliptical Dirac cone of the $y-z$ surface. This prefactor implies that
the Fermi velocity in $z$ direction is different from the Fermi
velocity in $y$ direction. On the $x-y$ surface, on the
other hand, the Dirac cone is circular and the Fermi velocities in $x$ and $y$ directions are identical.

Since superconductivity couples the electron and hole wave functions,
we write the surface states in the Nambu basis $\Psi = (\psi_{\uparrow},
\psi_{\downarrow}, \psi^{\dagger}_{\downarrow},
-\psi^{\dagger}_{\uparrow})^T$. The Hamiltonian for the surface states is given by
\begin{equation}
\mathbf{H_{p}}= \begin{pmatrix} H^{p} -\mu + U + \mathbf{M} & \Delta_0 e^{i \phi} \\
\Delta_0 e^{-i \phi}& - H^{p} + \mu - U + \mathbf{M} \end{pmatrix},
\label{eq:hamiltonian}
\end{equation}
where $H^{p}(\varepsilon_0^{p},v_F^{p})$ for $p = xy,yz$ denotes
the respective surface Hamiltonian, $\mu$ the chemical potential, $U$ the electrostatic potential,
$\Delta_0$ the induced superconducting pairing gap, $\phi$ the superconducting
phase, and $\mathbf{M} = \mathbf{m} \cdot \boldsymbol{\sigma}$, where $\mathbf{m}= (m_x, m_y, m_z)$ denotes
an induced exchange field.

Figures \ref{fig:bulk}(a), (b) and (c) show the geometries of the SNS
junctions that we study. They are divided into three regions: regions
I and III denote topological insulator surfaces with induced
superconductivity, whereas region II denotes a normal conducting
topological insulator surface (the weak link). The superconducting planes are produced
by bringing the surface in contact with an $s$-wave superconductor. The
proximity effect then induces effective $p$-wave superconductivity in
the surface states.\cite{Fu2008} It is assumed that there is an
electrostatic potential $U$ in the three regions which can be adjusted
independently by a gate voltage or doping. $U$ is measured from the chemical
potential in region II. The low energy states in region II are thus described by
Eq.~(\ref{eq:hamiltonian}) with $U = \Delta_0 = \mathbf{M} = 0$.  In
the superconducting regions I and III the potential is $U = -
U_0$. Furthermore, in the superconducting regions we have $\Delta_0
\neq 0$ and $\mathbf{M} = 0$.

Figures \ref{fig:bulk} (d) and (e) show the SFS junctions. The
ferromagnetic region II is established by placing a ferromagnetic
insulator on top of the topological insulator, which
induces an effective exchange coupling
due to proximity effect.\cite{Tanaka2009}
Consequently, the ferromagnetic region is described by
Eq.~(\ref{eq:hamiltonian}) with $\mathbf{M} \neq 0$ and $U = 0$.

We obtain $\Psi$ by solving the Bogoliubov-de-Gennes (BdG) equations $\mathbf{H_{p}} \Psi =
\varepsilon \Psi$. The wave functions in the superconducting regions are calculated in a
regime where $U_0 \gg |\mu-\varepsilon_0^p|, \varepsilon$. This means
that the Fermi wave length $\lambda'^p_F$ in the superconductor is
small, i.e., $\lambda'^p_F \ll \lambda_F^p,\xi $,
where $\lambda_F^p = \hbar v_F^p / \mu$ is the Fermi wave length in
the normal topological insulator surface and $\xi = \hbar v_F^p
/\Delta_0$ is the superconducting coherence length. We only consider
excitation energies smaller than the gap, $\varepsilon < \Delta_0$,
which implies that we only evaluate the Josephson current due to
Andreev bound states. Consequently, the momentum in $y$ direction
fulfills $|k_y| \le |(\mu - \varepsilon_0^p)/ \hbar
v_F^p|$, which allows us to simplify the wave functions. Furthermore,
we assume $U_0 +\mu - \varepsilon_0^{p} \gg
\Delta_0,\varepsilon$. Then, the surface states in region I of all
the junctions considered in this manuscript are described by
\begin{align}
\Psi_{all,S}^{I\pm}(x,y) &=   ( e^{\mp i \beta},\mp i e^{\mp i
  \beta},e^{-i \phi_I},\mp i e^{-i \phi_I})^T
 e^{i k_y y \pm i k_x x + \kappa x},
 \end{align}
where
\begin{align}
\beta &= \arccos(\varepsilon/\Delta_0)\:,\nonumber\\
k_x &= \sqrt{\frac{(U_0 +\mu -\varepsilon_0^{xy})^2}{(\hbar
      v_F^{xy})^2}- k_y^2}\:,\nonumber\\
\kappa &= \frac{(U_0 + \mu -\varepsilon_0^{xy})
\Delta_0}{(\hbar v_F^{xy})^2 k_x} \sin(\beta)\:.
\end{align}
Since $\varepsilon < \Delta_0$, the solutions decay exponentially for $
x \rightarrow - \infty$. The $\pm$ signs distinguish between waves propagating in positive and negative $x$ direction.

Similarly, the surface states in the superconducting region III of the
planar and step junction must vanish as $x \rightarrow +\infty$,
resulting in
\begin{align}
 \Psi_{planar,step,S}^{III\pm}(x,y) &= ( e^{\pm i \beta},\mp i e^{\pm
   i \beta },e^{-i \phi},\mp i e^{-i \phi} )^T
e^{i k_y y \pm i k_x x'- \kappa x'},
\end{align}
where $x' = x-L$ in case of the planar junction and $x' = x$ for the step
junction. In contrast, for the region III of the edge junction we obtain,
\begin{align}
 \Psi_{edge,S}^{III\pm}(y,z) &= ( e^{\pm i \beta},\pm i e^{\pm i \beta
 },e^{-i \phi},\pm i e^{-i \phi} )^T
e^{i k_y y \pm i k_z (z-L)- \kappa (z-L)},
\end{align}
with
\begin{align}
\eta k_z &= \sqrt{\frac{(U_0 +\mu
      -\varepsilon_0^{yz})^2}{(\hbar v_F^{yz})^2}- k_y^2}\:,\nonumber\\
\kappa &= \frac{(U_0 + \mu -\varepsilon_0^{yz})
    \Delta_0}{(\hbar v_F^{yz}\eta)^2 k_z} \sin(\beta)\:.
\end{align}

Next, we discuss the wave functions in the normal conducting segments of the SNS junctions. The wave functions in region II ($0 < x < L $, $z = 0$) for the planar
junction are:
\begin{align}
 \Psi_{planar,e}^{N\pm}(x,y) &= ( 1 , \mp i e^{\pm i
   \alpha(\varepsilon)} ,0,0 )^T
e^{i k_y y \pm i k_x(\varepsilon) x}, \\
\Psi_{planar,h}^{N\pm}(x,y) &= (0 , 0 , 1 , \mp i e^{\pm i
   \alpha(-\varepsilon)})^T
e^{i k_y y \pm i k_x(-\varepsilon) x},
\end{align}
where $\sin[\alpha(\varepsilon)] = \frac{\hbar v_F^{xy}
k_y}{\varepsilon+\mu -\varepsilon_0^{xy} }$ and $k_x (\varepsilon) =
\sqrt{\frac{(\varepsilon+\mu - \varepsilon_0^{xy})^2}{(\hbar
v_F^{xy})^2}-k_y^2}$. Due to the vanishing pair potential in this region, we find two independent solutions describing particles and holes, respectively, denoted by subscripts $e$ and $h$.

The wave functions for the step and the edge junctions in region II ($0 < z < L $, $x = 0$) are:
\begin{align}
\Psi_{step,edge,e}^{N\pm}(y,z) &= (\mp i \cos[\alpha(\varepsilon)] , 1 +
\sin[\alpha(\varepsilon)])^T
e^{i k_y y \pm i k_z(\varepsilon) z},
\end{align}
and
\begin{align}
\Psi_{step,edge,h}^{N\pm}(y,z) &=  (\mp i \cos[\alpha(-\varepsilon)] , 1 +
\sin[\alpha(-\varepsilon)])^T
e^{i k_y y \pm i k_z(-\varepsilon) z},
\end{align}
with $ \sin[\alpha(\varepsilon)] = \frac{\hbar v_F^{yz} k_y}
{\varepsilon+\mu -\varepsilon_0^{yz} }$, $\eta k_z
(\varepsilon) = \sqrt{\frac{(\varepsilon+\mu -
    \varepsilon_0^{yz})^2}{(\hbar v_F^{yz})^2}-k_y^2}$.
The meaning of the angle $\alpha$ can be understood in terms of the
Andreev reflection: $\alpha(\varepsilon)$ is the angle of incidence of
the electron (in momentum space) incident from the normal region to the superconducting region and
$-\alpha(-\varepsilon)$ is the reflection angle of the retroreflected hole. 


Region II of the ferromagnetic planar junction is described by the wave functions:
\begin{align}
 \Psi_{planar,e}^{F\pm}(x,y) &= \psi_{planar,e}^{F\pm} e^{i k_y y + (i \frac{m_y}{\hbar v_F^{xy}} \pm \kappa_e) x}, \\
 \Psi_{planar,h}^{F\pm}(x,y) &= \psi_{planar,h}^{F\pm} e^{i k_y y + ( -i \frac{m_y}{\hbar v_F^{xy}} \pm \kappa_h) x}
\end{align}
where
\begin{align*}
  \psi_{planar,e}^{F\pm}&= \left(\frac{\mu - \varepsilon_0^{xy} +
      \varepsilon}{\hbar v_F^{xy}} + \frac{m_z}{\hbar v_F^{xy}}, k_y
    +\frac{m_x}{\hbar v_F^{xy}} \mp \kappa_e \right)^T ,\\
  \psi_{planar,h}^{F\pm}&= \left( \frac{\mu -\varepsilon_0^{xy} -
      \varepsilon}{\hbar v_F^{xy}} - \frac{m_z}{\hbar v_F^{xy}}, k_y
    -\frac{m_x}{\hbar v_F^{xy}} \mp \kappa_h\right)^T, \\
  \kappa_e&=\kappa(\varepsilon,m_x), \\
  \kappa_h &=  \kappa(-\varepsilon,-m_x),\\
  \kappa(\varepsilon,m_x) &= \frac{\sqrt{m_z^2 - (\mu
      -\varepsilon_0^{xy}+ \varepsilon)^2 + (k_y
      \hbar v_F^{xy} + m_x)^2}}{\hbar v_F^{xy}}.
\end{align*}
The electron states are exponentially decaying if the magnetization is such that $(k_y
\hbar v_F^{xy} + m_x)^2 > (\mu -\varepsilon_0^{xy}+ \varepsilon)^2 -
m_z^2$, otherwise the electron states are propagating. Similarly, the hole states are
exponentially decaying for $(k_y \hbar v_F^{xy} - m_x)^2 > (\mu -\varepsilon_0^{xy}-
\varepsilon)^2 - m_z^2$. When we consider states at low energy, such
that $\mu -\varepsilon_0^{xy} \gg \varepsilon$, then the
magnetization in $x-$direction (in direction of transport) is
responsible for the difference in the decay of electron and hole
states.

In the ferromagnetic step junction we have:
 \begin{align}
 \Psi_{step,e}^{F\pm}(y,z) &= \psi_{step,e}^{F\pm}e^{i k_y y + \left(-i \frac{m_y}{\hbar v_F^{yz}\eta} \pm \kappa_e\right) z},\\
 \Psi_{step,h}^{F\pm}(y,z) &=  \psi_{step,h}^{F\pm}e^{i k_y y+ \left( i
   \frac{m_y}{\hbar v_F^{yz} \eta} \pm \kappa_h\right) z},
\end{align}
with
\begin{align*}
  \psi_{step,e}^{F\pm}&= \left( \frac{\mu - \varepsilon_0^{yz} +
      \varepsilon}{\hbar v_F^{yz}} -k_y(\varepsilon) +
    \frac{m_z}{\hbar v_F^{yz}},\frac{m_x}{\hbar v_F^{yz}} \pm \kappa_e
  \right)^T ,\\
  \psi_{step,h}^{F\pm}&= \left( \frac{\mu -
      \varepsilon_0^{yz} - \varepsilon}{\hbar v_F^{yz}}
    -k_y(\varepsilon) - \frac{m_z}{\hbar v_F^{yz}},-\frac{m_x}{\hbar
      v_F^{yz}} \pm \kappa_h\right)^T ,\\
  \kappa_e &= \kappa(\varepsilon,-m_z), \\
  \kappa_h &=\kappa(-\varepsilon,m_z), \\
  \kappa(\varepsilon,m_z) &=\frac{ \sqrt{m_x^2 - (\mu
      -\varepsilon_0^{yz}+ \varepsilon)^2 + (k_y \hbar v_F^{yz} +
      m_z)^2}}{\hbar v_F^{yz} \eta}.
\end{align*}
Here, the magnetization in $z$ direction, i.e., in direction of
transport, leads to a different decay length for electrons and holes. In
both ferromagnetic planar and step junctions, the transverse
magnetization ($y$ direction) does not lead to any exponential decay along the $y$ direction.

\subsection{\label{sec:Bc}Boundary conditions}
In a TR invariant system an interface between a superconducting and a normal conducting region can be described by a
single parameter $\gamma_k$ which determines the scattering at the
interface $k = 1,2$. \cite{Sen2012} Since the SNS junctions are
described by TR-invariant Hamiltonians, we derived such a boundary
condition similar to Ref.~[\onlinecite{Sen2012}] for our SNS
junctions. This finally leads to
the following boundary conditions for the planar junction at the
interface between region I and II:
\begin{align}
&[a^+_{e,h}\Psi_{planar,(e,h)}^{N +}(x,y) + a^-_{e,h} \Psi_{planar,(e,h)}^{N
  -}(x,y)]|_{x \rightarrow 0^+} = \\ & e^{- i \gamma_1 \sigma_y
}[\alpha^+\Psi_{all,S(e,h)}^{I+}(x,y) + \alpha^-
\Psi_{all,S(e,h)}^{I-}(x,y)]|_{x \rightarrow 0^-},\nonumber
\end{align}
where $e^{- i \gamma_1 \sigma_y }$ denotes the phase factor due to scattering
at the interface and  $a^{\pm}_{e,h}$ are the amplitudes of the electron
and hole wave functions propagating in $\pm x$ direction. In the
superconducting surface, on the contrary, the electron and hole wave
functions have the same amplitudes $\alpha^{\pm}$, as they are coupled via the BdG
equations. For the interface between region II and III we get similar
equations:
\begin{align}
&[\beta^+\Psi_{planar,S(e,h)}^{III+}(x,y) + \beta^-
\Psi_{planar,S(e,h)}^{III-}(x,y)]|_{x \rightarrow L^+} = \\ & e^{- i \gamma_2
  \sigma_y } [a^+_{e,h}\Psi_{planar,(e,h)}^{N +}(x,y) + a^-_{e,h} \Psi_{planar,(e,h)}^{N
  -}(x,y)]|_{x \rightarrow L^-},\nonumber
\end{align}
with the amplitudes $\beta^{\pm}$ for the superconducting wave
functions propagating in $\pm x-$direction.

For the step and edge junctions the interface between region I and II leads to the boundary condition
\begin{align}
&[a^+_{e,h}\Psi_{step,edge,(e,h)}^{N +}(y,z) + a^-_{e,h} \Psi_{step,edge,(e,h)}^{N
  -}(y,z)]|_{z \rightarrow 0^+} =\\ &i \sqrt{\frac{v_F^{xy}}{v_F^{yz}\eta}}e^{- i \gamma_1 \sigma_y
}[\alpha^+\Psi_{all,S(e,h)}^{I+}(x,y) + \alpha^-
\Psi_{all,S(e,h)}^{I-}(x,y)]|_{x \rightarrow 0^-}.\nonumber
\end{align}
The interface between region II and III of the step junctions is described by
\begin{align}
&[\beta^+\Psi_{step,S(e,h)}^{III+}(x,y) + \beta^-
\Psi_{step,S(e,h)}^{III-}(x,y)]|_{x \rightarrow 0^+} =\\& i
\sqrt{\frac{v_F^{yz} \eta}{v_F^{xy}}}e^{- i \gamma_2 \sigma_y }
[a^+_{e,h}\Psi_{step,(e,h)}^{N +}(y,z) + a^-_{e,h} \Psi_{step,(e,h)}^{N -}(y,z)]|_{z
  \rightarrow L^-},\nonumber
\end{align}
and of the edge junction by
\begin{align}
&[\beta^+\Psi_{edge,S(e,h)}^{III+}(x,y) + \beta^-
\Psi_{edge,S(e,h)}^{III-}(x,y)]|_{z \rightarrow L^+} = \\&e^{- i \gamma_2
  \sigma_y } [a^+_{e,h}\Psi_{edge,(e,h)}^{N +}(y,z) + a^-_{e,h} \Psi_{edge,(e,h)}^{N
  -}(y,z)]|_{z \rightarrow L^-}.\nonumber
\end{align}
In contrast to the planar junction, the Fermi velocities in the
superconducting and normal regions are different, and hence appear in the boundary conditions.

The boundary conditions yield eight equations and contain eight variables
($a^{\pm}_{e,h}$, $\alpha^{\pm}$ and $\beta^{\pm}$) and two parameters
$\gamma_1$ and $\gamma_2$ for the scattering at the first and the
second interface respectively. They can be written in a matrix
representation: $A \cdot (a^{+}_{e},a^{-}_{e},
a^{+}_{h},a^{-}_{h}, \alpha^{+} , \alpha^{-}, \beta^{+}, \beta^{-})^T
= 0.$ Nontrivial solutions exist if $\det(A)$ vanishes, so solving $\det(A) = 0$ as a function of $\varepsilon$ gives access to the bound state spectrum.
We include the phase difference
$\phi$ of the two superconducting regions by assuming that
the phase of region
I is $\phi/2$ and that of region III is $-\phi/2$.

The boundary conditions for the SFS type setups can be determined similarly. Since the proximity-induced ferromagnetism breaks TR symmetry, we used
for simplicity the continuity of the wave functions as boundary
condition for the ferromagnetic junctions.

\section{\label{sec:Results}Results and discussion}
\subsection{\label{sec:SNS}SNS junctions}
In this section we restrict ourselves to the step junction and the edge junction, since the
solutions for the planar junction can be obtained by a change of variables
(indices $yz \rightarrow xy$, $k_z \rightarrow k_x$, $\eta = A_1/A_2
\rightarrow 1$) from that of the step junction. Our results for the planar junction
are in agreement with similar calculations for a planar graphene SNS junctions.~\cite{Titov2006}

To calculate the Josephson current, a finite width $W$ is introduced
to quantize the transverse wave vectors in region II, $k_y \rightarrow
k_{yn}$ ($n = 0, 1,2, \ldots$). Denoting by $\rho_n(\varepsilon, \phi)$ the
density of states of mode $n$, the Josephson current at zero
temperature is given by
\begin{align}
J(\phi) = - \frac{2 e}{\hbar} \frac{d}{d \phi}\int_0^\infty
d\varepsilon \sum_{n = 0}^{\infty} \rho_n(\varepsilon, \phi)\varepsilon.
\end{align}
Using ``infinite mass'' boundary conditions \cite{Berry1987} at $y
= 0$ and $y=W$, the momentum is quantized to the values $k_{yn}= (n + 1/2) \pi /W$. This quantizes
$k_{zn}$ and $\eta k_{zn} = \sqrt{\left(\frac{\mu -
      \varepsilon_0^{yz}}{\hbar v_F^{yz}}\right)^2-k_{yn}^2}$, which
means the lowest $N(\mu - \varepsilon_0^{yz}) = \left(\frac{\mu -
    \varepsilon_0^{yz}}{\hbar v_F^{yz}}\right) \frac{W}{\pi} $ modes
are propagating as $k_{zn}$ is real, while the higher modes are
evanescent, since for these modes $k_{zn}$ is imaginary. The analysis
of the Josephson current is done in the short-junction regime where
the length $L$ of the normal region is small relative to the
superconducting coherence length $\xi$ and $L \ll W$. This requires $
\Delta_0 \ll \hbar v_F^{yz} /L$ making $\alpha(-\varepsilon) \approx
\alpha(\varepsilon) \approx \alpha(0) =:\alpha$ and
$k_z(-\varepsilon)\approx k_z(\varepsilon)\approx k_z(0)=:k_{zn}$ a
good approximation. The solution is a single bound state per mode:
\begin{align}
 \varepsilon_n(\phi) &= \Delta_0 \sqrt{1 - \tau_n \sin^2(\phi/2)},\label{energymodesstep} \\
 \tau_n &= \frac{\left(\eta k_{zn}\right)^2}{\left(\eta k_{zn}\right)^2\cos^2(k_{zn} L) + \left(\frac{\mu - \varepsilon_0^{yz}}{\hbar v_F^{yz}}\right)^2 \sin^2(k_{zn}L)}. \notag
\end{align}
Here, $\tau_n$ can be interpreted as the transmission probability of the topological insulator
surface sandwiched between two topological
superconducting surfaces.

\begin{figure}[t]
\includegraphics[width = 0.99\columnwidth]{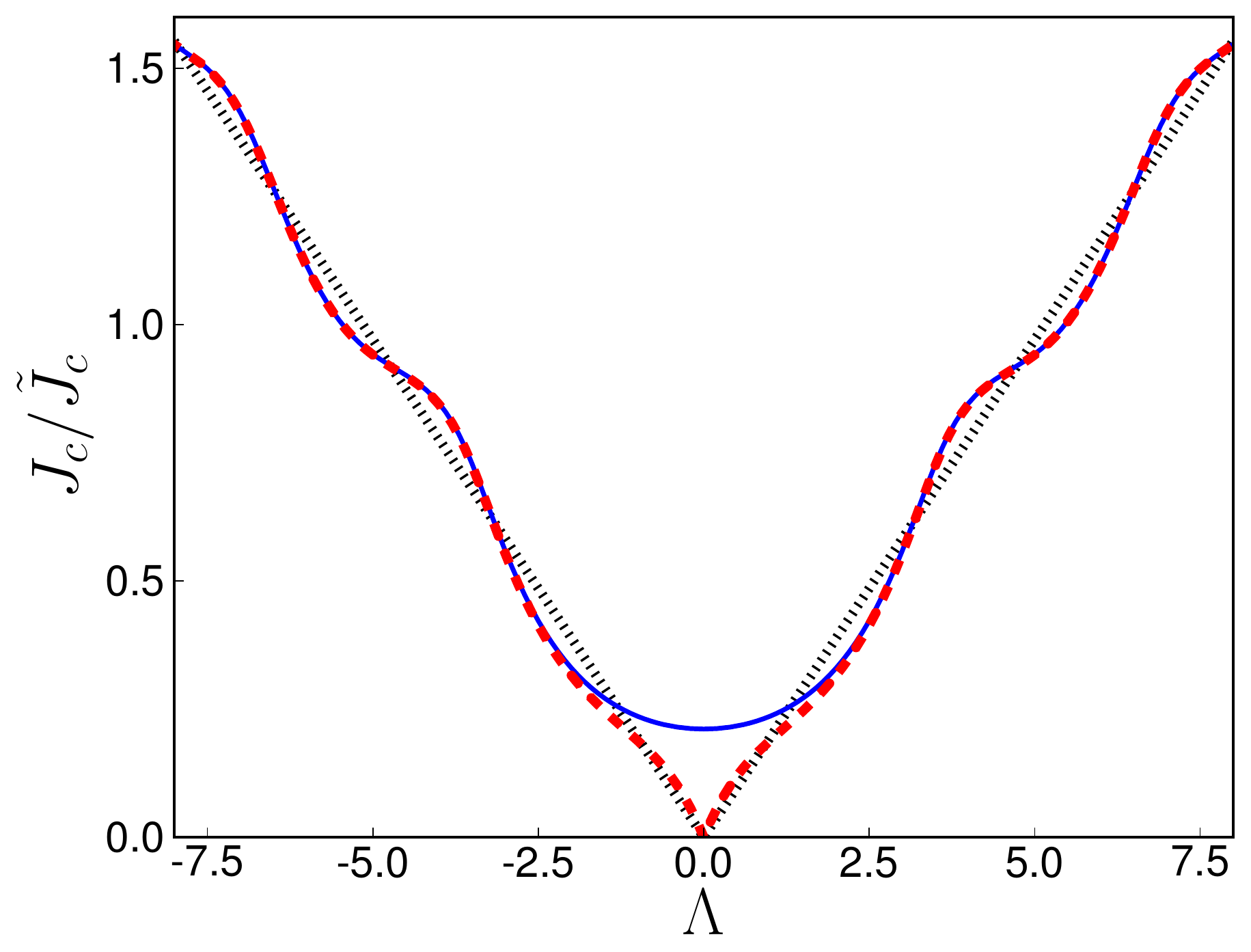}
\caption{\label{step}(Color online) Critical Josephson current $J_c$ (in units of $\tilde{J_c}$) of a step
  junction as a function of the
  chemical potential measured from the Dirac point energy $\Lambda$, see Eq.~(\ref{eq:defLambda}). The dashed red line shows the contribution of the
  propagating waves only, and the dotted black line is the asymptote for $|\Lambda| \gg 1$.}
\end{figure}

By using $\rho_n(\varepsilon, \phi) = \delta[\varepsilon -
\varepsilon_n(\phi)]$ the supercurrent due to the discrete spectrum
becomes
\begin{align}
 J(\phi)& = \frac{ e \Delta_0}{2 \hbar} \sum_{n = 0}^{\infty} \frac{\tau_n \sin(\phi)}{ \sqrt{1 - \tau_n \sin^2(\phi/2)}} \label{currentintegral1}\\
 & = \frac{ e \Delta_0}{\hbar} \frac{W}{2 \pi} \int_{0}^{\infty} \frac{\tau_n \sin(\phi)}{ \sqrt{1 - \tau_n \sin^2(\phi/2)}} dk_{yn} \nonumber \\
 & = \tilde{J_c}\frac{L}{2 \pi \eta} \int_{0}^{\infty} \frac{\tau_n \sin(\phi)}{ \sqrt{1 - \tau_n \sin^2(\phi/2)}} dk_{yn} \nonumber 
\end{align}
where 
\begin{align}\label{eq:defJc}
    \tilde{J_c} = \frac{ e \Delta_0}{\hbar} \frac{W}{L} \eta
\end{align}
and
the summation over $n$ has been replaced by an integration (since $L \ll W$). By maximizing the current with respect to $\phi$, the
critical Josephson current can be calculated, see
Fig.~\ref{step} (blue solid line).

In Fig.~\ref{step} we plot the normalized critical current $J_c/\tilde{J_c}$ as a function of the rescaled energy,
\begin{align}\label{eq:defLambda}
\Lambda=\frac{\mu-
  \varepsilon_0^{yz}}{\hbar v_F^{yz}} \left(\frac{L
  }{\eta}\right)    
\end{align}
Obviously, the critical current is dependent on $\eta = A_1/A_2$. Furthermore, we find a finite critical Josephson
current for chemical potential at the Dirac point energy ($\Lambda= 0$). By comparing this to the critical current due to propagating
waves only (Fig.~\ref{step}, dashed red line) we find that this
finite current appears due to evanescent waves, i.e, due to
imaginary $k_z$. This critical current at the Dirac point energy can
be tuned by the fraction $\eta = A_1/A_2$: the larger $\eta$, the larger
the critical current. The critical current for $\big|\Lambda\big|\gg 1$ follows the asymptote shown by
the dotted black line in
Fig.~\ref{step}. The oscillations in the critical current can be
considered as a negligible deviation in this limit.

Performing the same calculations for an edge junction yields
\begin{align}
 \varepsilon_n &= \Delta_0 \sqrt{(1 - \tau_n \sin( \phi / 2)^2)}, \\
\tau_n &= - \frac{1}{\tan(\alpha)^2 \sin(k_{zn} L)^2 }. \notag
\end{align}
Naturally, the result $\tau_n < 0$ corresponds to $\varepsilon_n > \Delta_0$,
implying the absence of Andreev bound states. The formation of Andreev
bound states in the central region requires the presence of electrons
with opposite spins in regions I and III. Due to spin momentum locking
in the topological insulator and because the spins in region I and III
lie in different planes, the formation of Andreev bound states is
prohibited. The only possibility would be $k_y = 0$, in which case the spin is along
the $y$ direction. However, this is not allowed due to the boundary conditions:
$k_{yn} = (n + 1/2) \pi/W > 0$. Thus, the contribution of the Andreev
bound states to the Josephson current vanishes in these edge
junctions.

\subsection{\label{sec:SFS}SFS junctions}
Again we focus on the step junction, since we can get the
solutions for the planar junction by a change of variables (indices
$yz \rightarrow xy$, $\eta = A_1/A_2 \rightarrow 1$) and of the
magnetization direction. First, we examine a junction with perpendicular magnetization and
later on we will analyze the effects of the magnetization in all
directions for the case, where the chemical potential is at the Dirac point
energy.

\subsubsection{Perpendicular magnetization ($m_z=0$ and $m_y =0$)}
\begin{figure}[t]
\includegraphics[width = 0.99\columnwidth]{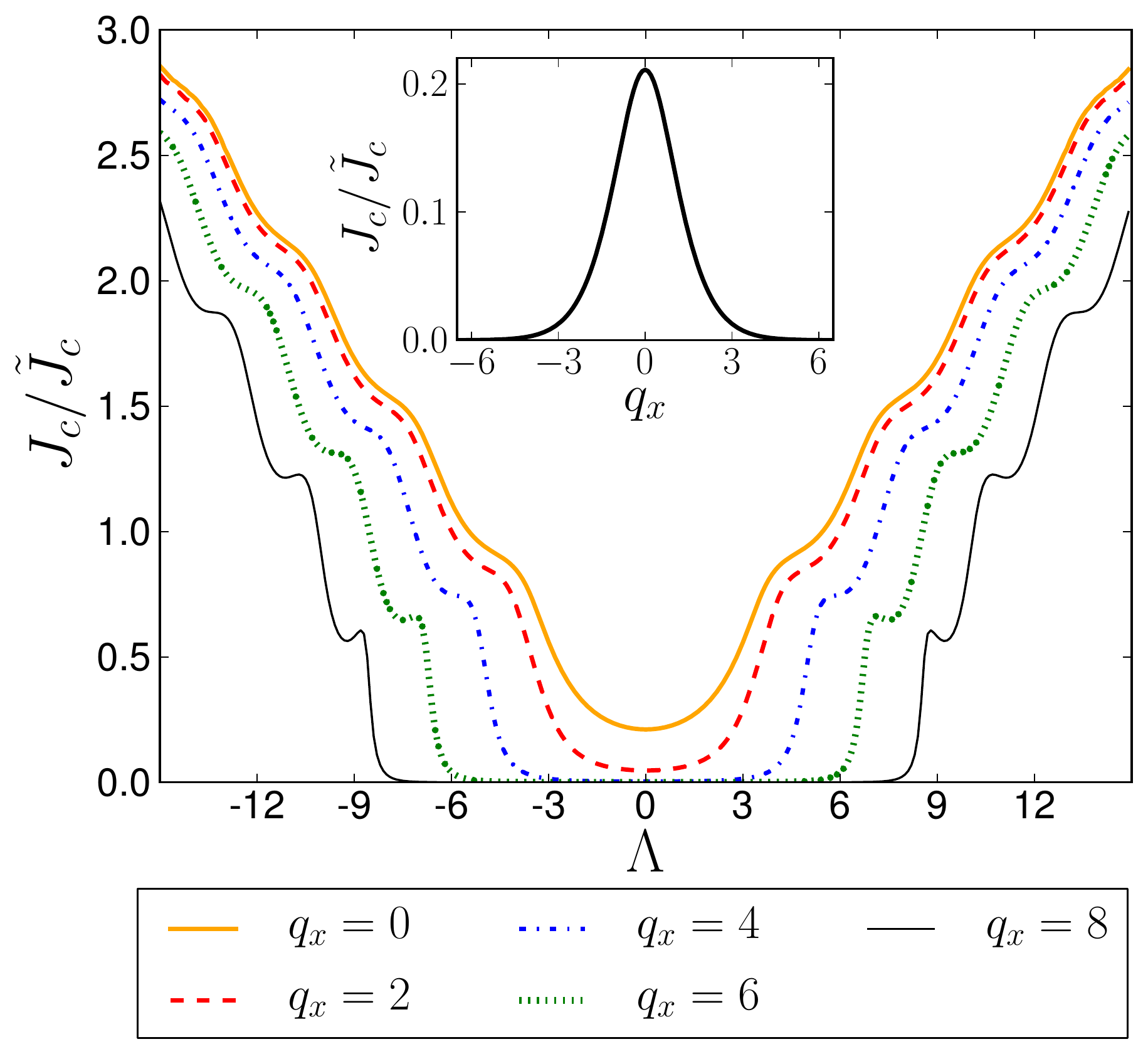}
\caption{(Color online) Critical Josephson current $J_c$ (in units of $\tilde{J_c}$) of a
   ferromagnetic step junction as a function of $\Lambda$, see Eq.~(\ref{eq:defLambda}), for different values of $q_x = m_x L/(\hbar
     v_F^{yz} \eta)$. Inset: $J_c$ as a function of $q_x$ at the Dirac point
   ($\Lambda = 0$). }
\label{fig:Jcferrostepgeneral}
\end{figure}

In the low-energy regime, i.e., for ($|\mu - \varepsilon_0^{yz}| \gg
\varepsilon$) we can use $\kappa(-\varepsilon,-m_z) \approx
\kappa(\varepsilon,m_z)\approx \frac{1}{\hbar
  v_F^{yz} \eta}\sqrt{m_x^2 - (\mu
  -\varepsilon_0^{yz})^2 + (k_y \hbar v_F^{yz})^2} =: \kappa$ resulting in the following energy:
 \begin{equation}
 \varepsilon = \Delta_0 \sqrt{\frac{\kappa^2 + k_y^2 \sinh^2(\kappa L)-\kappa^2 \sin^2(\phi/2)}{\kappa^2 + (k_y^2 + m_x^2)\sinh^2(\kappa L)}}.
 \end{equation}
By introducing a finite width $W$ which quantizes $k_y \rightarrow
k_{yn}$ ($n = 0, 1,2, \ldots$) and by using the ``infinite mass''
boundary condition the supercurrent due to the discrete spectrum is
\begin{align}
 J(\phi) &= 
 \frac{ e \Delta_0}{\hbar} \frac{W}{2 \pi} \int_{0}^{\infty}
 \frac{\kappa_n^2 \sin(\phi)\varepsilon^{-1}_n(\phi)}{\kappa_n^2 +
   (k_{yn}^2 + m_x^2)\sinh^2(\kappa_n L)}
 dk_{yn} .\nonumber
\end{align}
The critical current can be calculated for different values of $q_x =
\frac{m_x}{\hbar v_F^{yz}} \frac{L}{\eta}$, see
Fig.~\ref{fig:Jcferrostepgeneral}. For $q_x = 0$ the solutions of the
normal step junction are recovered. The calculation shows that for $\Lambda
= 0$ and $q_x = 0$ we get the critical current
$J_c = 0.21 \tilde{J_c}$. 

Figure~\ref{fig:Jcferrostepgeneral} shows that the magnetization
$q_x$ can be used to tune the critical Josephson current. The stronger
the magnetization, the larger the chemical potential needs to be, to
result in a finite current. For large magnetization the finite
Josephson current at the Dirac point ($\Lambda = 0$) vanishes. This
can be seen in the inset of Fig.~\ref{fig:Jcferrostepgeneral} which
shows the dependence of $J_c$ on $q_x$ at $\Lambda = 0$.

At larger values of $\Lambda$ and $q_x$ we get a non-monotonic
behavior, as it can be seen for $q_x = 8$ in
Fig.~\ref{fig:Jcferrostepgeneral}. We compare the values of $\Lambda$,
where this non-monotonic behavior arises, to the scaled number of
modes $N' = N L / (W \eta) =
\sqrt{\Lambda^2-q_x^2}/\pi$. When $\Lambda$ is
small enough such that the curve is still monotonic, we find that $N' < 1$. We
increase $\Lambda$ and just before the first local maximum of $J_c$,
$N'$ exceeds $1$. Similarly, just before the second local maximum $N$ increases
further by 1.

\begin{figure}[t]
\includegraphics[width = 0.99\columnwidth]{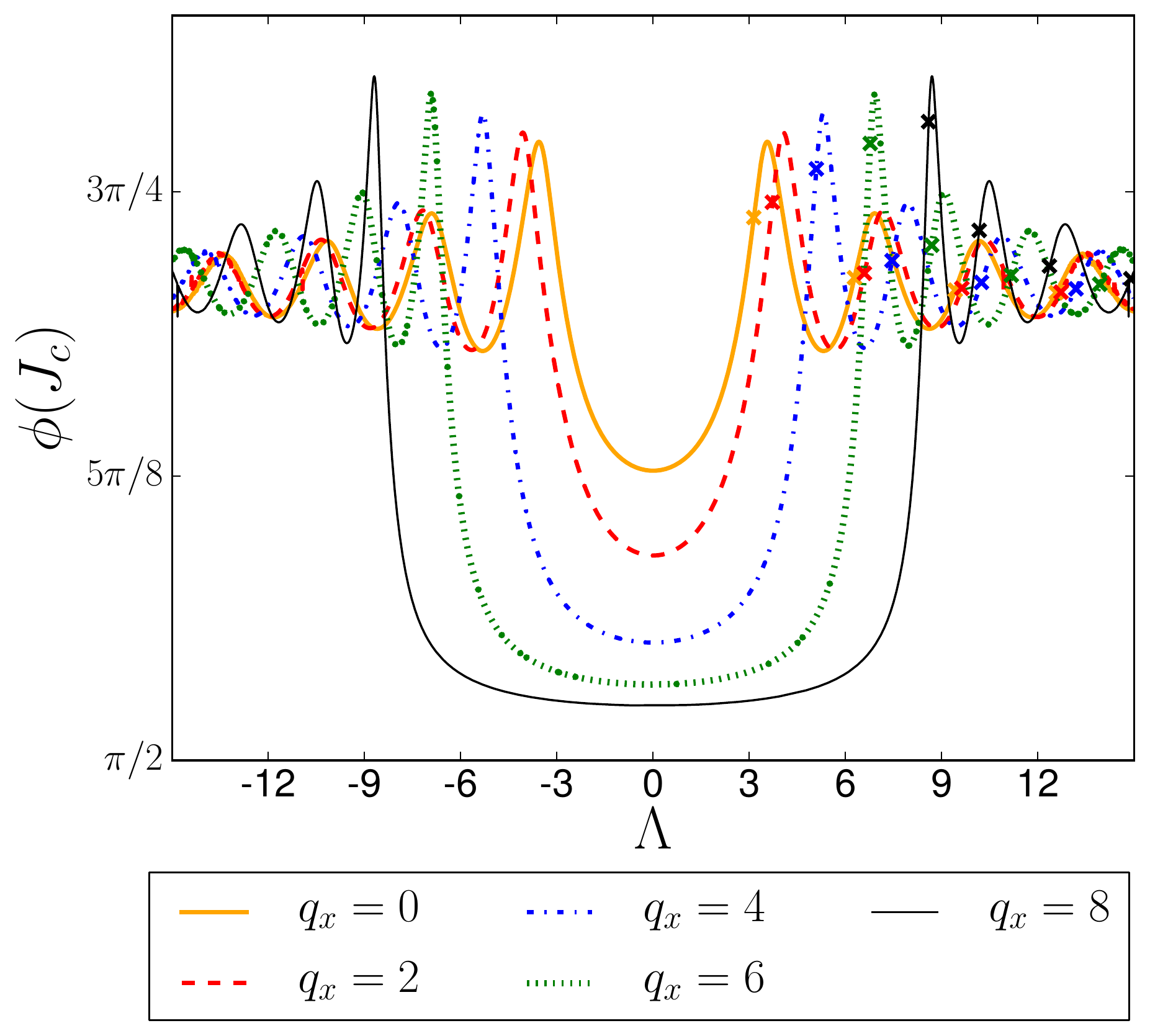}
\caption{(Color online) Superconducting phase difference $\phi (J_c)$ (which
  maximizes the current $J$) as a function of $\Lambda$, see Eq.~(\ref{eq:defLambda}), for different values of $q_x = m_x L/(\hbar
    v_F^{yz} \eta)$. The crosses indicate the values of
   $\Lambda$ at
  which the scaled number of modes $N' = N L / (W \eta)= \sqrt{\Lambda^2-q_x^2)}/\pi$ takes an integer
  value.}
\label{fig:Jcphi}
\end{figure}

When plotting the superconducting phase difference $\phi (J_c)$ which
maximizes the current $J$ against $\Lambda$, we find that this phase
difference oscillates and the amplitude of the oscillations increase with $q_x$. This behavior is
shown in Fig.~\ref{fig:Jcphi}. The crosses indicate the values of
$\Lambda$ where $N'$ becomes an integer. Again we see that by
increasing $\Lambda$ the next integer value of $N'$ is achieved just
before the maximum of $\phi(J_c)$. This correlation suggests that this
non-monotonic behavior arises due to quantum interference of the new
additional propagating mode and the already existing ones.

\subsubsection{Chemical potential at the Dirac point energy ($\mu = \varepsilon_0^{yz}$)}

\begin{figure}[t]
\includegraphics[width = 0.99\columnwidth]{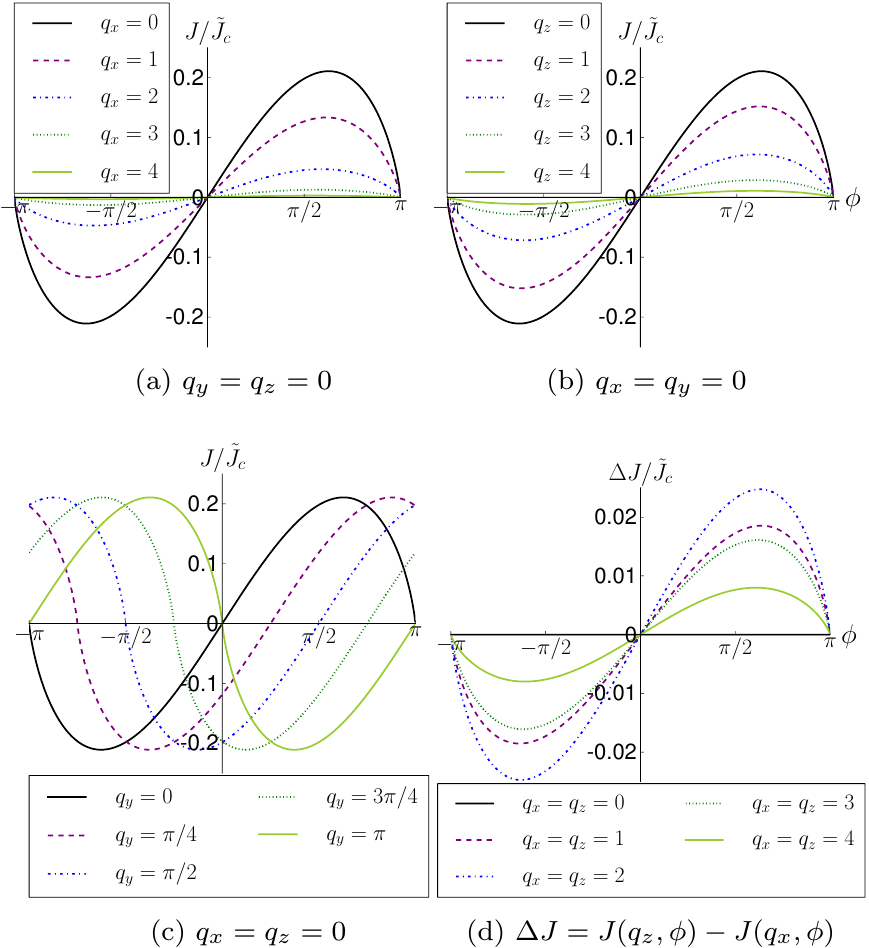}
\caption{(Color online) Josephson current $J$ (in units of $\tilde{J_c}$) for the ferromagnetic step junction as a function of
  $\phi$ for different values of the magnetization parametrized by
    $q_{x,y,z} = m_{x,y,z} L/(\hbar v_F^{yz} \eta)$. The black line in (a), (b) and (c)
  corresponds to $q_x=q_y=q_z=0$. (d) shows the difference between the Josephson
  current of junctions with magnetization in $x-$ and $z-$direction
  ($\Delta J = J(q_z,\phi) - J(q_x,\phi)$).}
\label{fcurrentmu0}
\end{figure}

In the limit $\mu = \varepsilon_0^{yz}$, it becomes possible to analytically examine arbitrary directions of the magnetization.
If we assume $\sqrt{m_x^2 +m_z^2} \gg \varepsilon$
then $ \kappa(\varepsilon,-m_z) \approx \kappa(0,-m_z) =:
\kappa(-m_z)$ and $\kappa(-\varepsilon,m_z) \approx
\kappa(0,m_z)=:\kappa(m_z)$. Again we calculate the energy and find
\begin{align}
\varepsilon &= \frac{\Delta_0}{\sqrt{2}}\sqrt{1+ \frac{1}{\eta^2} \left(k_y^2 - \frac{m_x^2}{(\hbar v_F^{yz})^2}-\frac{m_z^2}{(\hbar v_F^{yz})^2} \right)f_1 + f_2(\phi)}, \label{eq:energy} \\
 f_1 &= \frac{\sinh(\kappa(m_z)L)\sinh(\kappa(-m_z)L)}{\kappa(m_z)\kappa(-m_z)\cosh[\kappa(m_z)L]\cosh[\kappa(-m_z)L]}, \nonumber \\
 f_2(\phi) &= \frac{\cos(2 L \frac{m_y}{\hbar v_F^{yz} \eta}
   - \phi)}{\cosh[\kappa(m_z)L]\cosh[\kappa(-m_z)L]}. \nonumber
\end{align}
The energy fulfills $ \varepsilon \leq \Delta_0$. The current is
calculated in the same way as before.  We can see from
Eq.~(\ref{eq:energy}) and in Fig.~\ref{fcurrentmu0}(c) that the
magnetization $\frac{m_y}{\hbar v_F^{yz} \eta}$ leads to a
phase shift in the Josephson current and thus does not influence the
critical current. Furthermore, we observe from
Fig.~\ref{fcurrentmu0}, that a magnetization in the
$x$ direction (out of plane) suppresses the current
more strongly than the magnetization in the $z$ direction (in plane and in the direction of
transport).

\section{\label{sec:Conclusion}Conclusion}
We have conducted a detailed study of the Josephson effect on the
surface of a topological insulator, using Bi$_2$Se$_3$ as a model
system.  The symmetries of the bulk crystal structure give rise to
different Fermi velocities along the rotation symmetry axis and in the
direction perpendicular to it. This manifests itself in a scaling
factor in the critical Josephson current of the step
junction when compared to the planar junction. This scaling appears in
both normal and ferromagnetic topological insulator junctions.
Interestingly, the contribution to the Josephson current from Andreev
bound states vanishes for the edge junction. This
suppression can be explained in terms of spin momentum locking, which
prohibits the formation of Andreev bound states in the central region.
In the ferromagnetic topological insulator step junction, we find that
the critical Josephson current is suppressed for an out of plane magnetization
as well as for a magnetization, which is in plane and in the direction of transport.
A magnetization along the junction and perpendicular to the direction of transport leads
to a finite Josephson current even when the phase difference of the
superconductors is zero. Finally, we have obtained a non-monotonic
critical Josephson current when the perpendicular magnetization and
the chemical potential are sufficiently large, which was explained in
terms of quantum interference of multiple modes in the junction. An
experimental verification of this behavior (as shown in
Fig.~\ref{fig:Jcferrostepgeneral}) could provide valuable insights
into the transport mechanisms in these junctions.

\begin{acknowledgments}

We would like to acknowledge stimulating discussions with Julia
Meyer.  This work was financially supported by the Swiss SNF and the
NCCR Quantum Science and Technology.

\end{acknowledgments}


\end{document}